\documentstyle[aasms4,psfig]{article}

\received{} \accepted{}

\lefthead{Du et al.}
\righthead{The metallicity distribution of
F/G dwarfs with BATC survey}

\begin{document}

{\title{The metallicity distribution of F/G dwarfs derived from
BATC survey data}

\author{
 Cui-hua Du\altaffilmark{1,2},
 Xu Zhou\altaffilmark{1},
 Jun Ma\altaffilmark{1},
 Jian-rong Shi\altaffilmark{1},
 Alfred Bing-Chih Chen\altaffilmark{3},\\
 Zhao-ji Jiang\altaffilmark{1},
 Jian-sheng Chen\altaffilmark{1}
  }

\altaffiltext{1}{National Astronomical Observatories, Chinese
Academy of Sciences, Beijing, 100012, P. R. China;
dch@vega.bac.pku.edu.cn} \altaffiltext{2}{College of Physical
Sciences, Graduate School of the Chinese Academy of Sciences,
Beijing, 100012, P. R. China} \altaffiltext{3}{Department of
Physics, National Cheng Kung University, Taiwan 70148, Taiwan }

\authoremail{dch@vega.bac.pku.edu.cn}

\begin{abstract}
Based on synthetic flux spectra calculated from theoretical
atmospheric models, a calibration of temperature and metallicity
for the dwarfs observed in the Beijing-Arizona-Taiwan-Connecticut
(BATC) multicolor photometric system is presented in this paper.
According to this calibration, stellar effective temperatures can
be obtained from some temperature--sensitive color indices. The
sample stars have colors and magnitudes in the ranges
$0.1<d-i<0.9$ and $14.0<i<20.5$. The photometric metallicities for
these sample stars can be derived by fitting SEDs. We determine
the average stellar metallicity as a function of distance from the
Galactic plane. The metallicity gradient is found to be
d[Fe/H]/d$z=-0.37\pm0.1$ dex/kpc for $z<4$ kpc and
d[Fe/H]/d$z=-0.06\pm0.09$~dex/kpc between 5 and 15 kpc. These
results can be explained in terms of different contributions in
density distribution for Galactic models `thin disk', `thick disk'
and `halo' components. However, for the gradient in $z>5$ kpc,  it
could not be interpreted according to the different contributions
from the three components because of the large uncertainty. So it
is possible that there is little or no gradient for $z>5$ kpc. The
overall distribution shows a metallicity gradient
d[Fe/H]/d$z=-0.17\pm0.04$~dex/kpc for $z<15$ kpc.
\end{abstract}


\section{INTRODUCTION}

The Galaxy is unique in offering the possibility of determining
directly the three-dimensional distributions of luminous mass and
chemical abundances. Combination of these results allows a
detailed investigation of the dominant physical processes that
occurred during the formation and early evolution of the Galaxy
(Gilmore \& Wyse 1985). Over the past decade considerable efforts
have been undertaken to gain information about the structure and
formation of the Galaxy.

Of crucial importance in the structure and formation of the Galaxy
is the existence of a thick disk component. Ever since it was
revealed that the Galactic disk contains two distinct stellar
populations, the origin and nature have been discussed by an
number of investigators. The thick disk population was evident in
the data of Hartkopf \& Yoss (1982) and clearly explained by
Gilmore \& Reid (1983). While subsequent investigations have
determined the overall kinematic and chemical properties of the
thick disk rather well (Gilmore \& Wyse 1985; Sandage \& Fouts
1987; Majewski 1992; Reid et al.~1993; Robin et al.~1996), the
relation of the thick disk to the halo population and to the thin
disk as well as its evolutionary history are still poorly known
(Gilmore 1989).  Also, the formation of the thick disk component
is also an open question. A number of models for the formation of
the thick disk have been put forward since the confirmation of its
existence. Sandage (1990) proposed that the thick disk was formed
by dissipative pressure-supported collapse in the early history of
the Galaxy, producing a kinematic and chemical gradient as the gas
was settling into the galactic plane. Other scenarios of the thick
disk formation, including that in which the thick disk is the
result of an accretion induced heating of the thin disk, were
discussed by Majewski (1993). In order to study the formation
process of the thick disk in more detail, additional information
such as the metallicity gradient is also needed. If the detailed
information becomes available, it is possible to provide clues not
only for the formation history of the thick disk, but also for the
Galaxy as a whole.

Numerous surveys along the Galactic plane have been used to
investigate the existence and size of the galactic radial
abundance gradient in the disk. A wide variety of objects have
been used to determine this gradient, from radio and optical
observation of HII regions, disk stars (Neese \& Yoss 1988),
planetary nebulae (Shaver et al. 1983; Pasquali \& Perinotto 1993;
Henry \& Worthey 1999; Hou et al. 2000; Maciel et al. 2003) and
open clusters (Friel 1995; Chen et al.~2003). The existence of a
radial gradient in the Galaxy is now well established. An average
gradient of about $-0.06$ dex kpc$^{-1}$ is observed in the
Galactic disk for most of the elements (Chen et al, 2003). On the
other hand, there is considerable disagreement about whether there
is a vertical metallicity gradient among field and/or open cluster
stars of the Galactic disk. Its extent and reasons are not well
understood either, yet potentially it must be a powerful clue to
the Galactic formation.  The amplitude of any vertical gradient of
stellar properties allows one to place constraints on the
existence of discrete stellar subpopulation and/or distinct
components of the Galaxy (Sandage 1981).

The BATC multicolor photometric survey accumulated a large data
base which is very useful for studying the structure and formation
of the main components of the Galaxy. Du et al.~(2003) provided
some information on the density distribution of the main
components of the Galaxy, which can present constraints on the
parameters of models of the Galactic structure. Here, we shall use
F and G dwarfs from the BATC survey data to provide the
metal-abundance information. The main sequence lifetime of F and G
type stars is longer than the age of the Galaxy, and hence the
chemical abundance distribution function of such stars provides an
integrated record of the chemical-enrichment history (Chen et
al.~2000). Many general
trends have been discovered during the past decades.
With the new improved observation and improved
knowledge regarding galaxy formation, it becomes possible to
further discuss the metallicity gradient in the Galaxy.

In this paper, we attempt to study the metallicity gradient of the
Milky Way galaxy using the BATC photometric survey data. The
outlines of this paper is as follows. The BATC photometric system
and data reduction are introduced briefly in Sect.\,2. In Sect.\,3
we describe the theoretical model atmosphere spectra and synthetic
photometry.  In Sect.\,4,  we present the BATC data used in this
analysis. The vertical metallicity gradient is discussed in
Sect.\,5. In Sect.\,6, an estimation for metallicity gradient
using a three-component model is made and the result is compared
with the observed metallicity gradient. Finally, in Sect.\,7 we
summarize our main conclusions in this study.
\section{BATC PHOTOMETRIC SYSTEM AND DATA REDUCTION}

The BATC program uses the 60/90 cm f/3 Schmidt telescope at the
Xinglong Station of the National Astronomical Observatories
(NAOC), with a $2048 \times 2048$ Ford CCD mounted at its focal
plane. The field of view of the CCD is $58^{\prime}$ $\times $ $
58^{\prime}$, with a pixel scale of
$1\arcsec{\mbox{}\hspace{-0.15cm}.}7$. There are 15
intermediate-band filters in the BATC filter system, which covers
an optical wavelength range from 3000 to 10000 {\AA}. Fig.\,1
shows the filters transmissions. The well-known advantage of using
these filters is that intermediate bandwidths can ignore color
terms for atmospheric extinction correction, as opposed to such a
necessity for broad band filters. The BATC magnitudes adopt the
monochromatic AB magnitudes as defined by Oke \& Gunn (1983). The
standard stars HD19445, HD84937, BD+262606 and BD+174708 (Oke \&
Gunn 1983) are observed for flux calibration in the BATC survey.
The detailed description of the BATC photometric system and flux
calibration of the standard stars can be found in Fan et
al.~(1996) and Zhou et al.~(2001, 2003).

The BATC survey images were reduced through standard procedures,
including bias subtraction, flat-fielding correction and flux
calibrations (Fan et al. 1996; Zhou et al. 2001; Zhou et al.
2003). After the basic corrections described above, the multiple
field images observed of each filter were combined by integer
pixel shifting, respectively. When combining the images, the
cosmic rays and bad pixels were corrected by comparison of
multiple images. The HST Guide star catalog (GSC) (Jenkner et al.
1990) was then used for coordinate determination. The final RMS
error in positions of GSC stars is about 0.5 arcsec. The
magnitudes of the point sources in the BATC fields are measured by
the photometric method of point spread function (PSF) fitting. Our
PSF magnitudes were obtained through an automatic data reduction
program PIPELINE 2, which  was developed based on Stetson's
DAOPHOT procedures (Stetson 1987). Finally,  at the completion of
photometry, the spectral energy distribution (SEDs) of all
measurable objects are obtained.

\begin{figure}
\figurenum{1} \centerline{\psfig{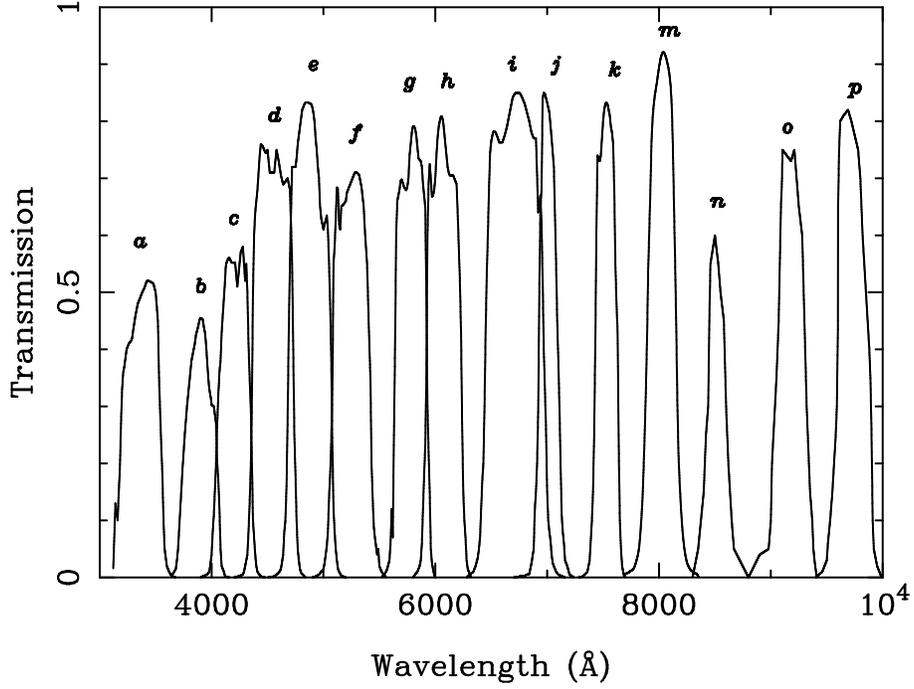}}
\caption{The transmission curves for the BATC multicolor filters.}
\end{figure}

\section{THEORETICAL MODEL AND CALIBRATION FOR TEMPERATURE AND METALLICITY}

\subsection{ Theoretical stellar library and synthetic photometry}
A homogeneous and complete stellar library can match
any ambitious goals imposed on a standard library. Lejeune et al.~(1997)
presented a hybrid library of synthetic stellar spectra.
The library
covers a wide range of stellar parameters: $T_{\rm eff}$=50,000 K to
2,000 K in intervals of 250 K,
log~$g$=$-1.02$ to 5.50 in main increments of 0.5,
and [M/H]=$-5.0$ to +1.0. For each model in the library,
a flux spectrum is given for the same set of 1221 wavelength points covering the range
9.1 to 160,000 nm, with a mean resolution of 20{\AA} in the visible.
The spectra are thus in a format which has proved to be adequate
for synthetic photometry of wide band and intermediate band systems.

Based on the theoretical library, we calculate synthetic colors of
the BATC system. Here, we synthesize colors for simulated stellar
spectra with $T_{\rm eff}$ and log~$g$ characteristic of F/G
dwarfs (log~$g$=4.0, 4.5 for dwarfs) and 19 values of metallicity
([M/H]=$-5.0$, $-4.5$, $-4.0$, $-3.5$, $-3.0$, $-2.5$, $-2.0$,
$-1.5$, $-1.0$, $-0.5$, $-0.3$, $-0.2$, $-0.1$, 0.0, +0.1, +0.2,
+0.3, +0.5 and +1.0), where [M/H] denotes metallicity relative to
hydrogen. The synthetic $i$th BATC filter magnitude can be
calculated with
\begin{equation}
 m = -2.5~{\rm log}\frac{\int{F_{\lambda}\phi_{i}({\lambda}){\rm d}\lambda}}
{\int{\phi_{i}({\lambda}){\rm d}\lambda}} - 48.60,
\end{equation}
where ${F_{\lambda}}$ is the flux per unit wavelength, $\phi_{i}$
is the transmission curve of the $i$th filter of the BATC filter
system (Fig.\,1). In Fig.\,2, as an example, we present the
two-color diagram based on $c$, $i$, $p$ filters. From this
figure, we can clearly see how the color indices vary as a
function of temperature and metallicity. In this two-color
diagram, the gravity log~$g$=4.5 corresponds roughly to dwarfs.
Four metallicities: [M/H]=$-5.0$, $-1.5$, 0.0 and +1.0 are
arbitrarily selected and the temperature range is 5000 K $\le$
$T_{\rm eff}$ $\le$ 8000 K. For clarity, the grid points are
connected with lines, illustrating isothermal and iso-metallicity
lines. Similar grids, of course, exist as well for other two-color
diagrams with different gravities and metallicities in the
library.

\begin{figure}
\figurenum{2} \centerline{\psfig{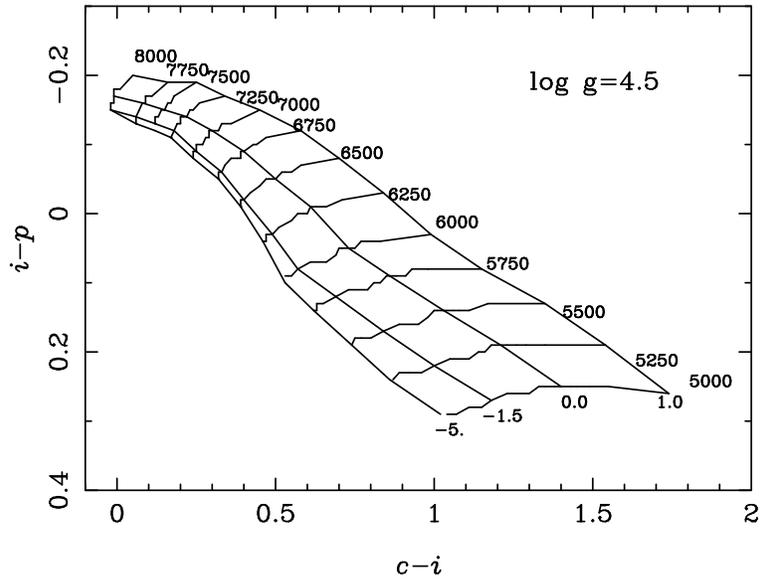}}
\caption{The $i-p$ versus $c-i$ diagram. A grid of constant-log~g
models with the metallicities: [M/H]=$-5.0$, $-1.5$, 0.0 and +1.0;
and with $T_{\rm eff}$ $\ge$ 5000 K are presented in the two-color
diagram. The grid points are connected with straight lines,
illustrating isotherm and iso-metallicity lines.}
\end{figure}

The bluer colors are sensitive to metallicity down to the lowest
observed metallicities because most of the line-blanketing from
heavy elements occurs in the shorter wavelength regions. In
contrast, the redder colors are primarily sensitive to temperature
index. The BATC $a$, $b$ bands contain the Balmer jump, a stellar
spectral feature which is sensitive to surface gravity. Since our
sample includes only F/G dwarfs, it conveys little gravity
information. Therefore, we shall focus on effective temperature
and metallicity in the following sections. It should be mentioned
that, although the metallicity or temperature derived from
synthetic photometry is not very accurate for a single star,
perhaps which can be distorted by a poor point, it is meaningful
by studying sample stars.

\subsection{Temperature and color index}
The stellar effective temperature can be derived by comparing
observed SEDs with theoretical ones. In general, a sensitive color
index can represent this physical quantity, such as $b-y$ in the
$ubvy$ system (Ardeberg et al.~1983). We would like to find such
color indices which are sensitive to the effective temperature in
the BATC filter system. Our study shows that color indices $d-n$,
$d-o$, $e-o$ and $e-p$ are very sensitive to temperature but are
relatively insensitive to log~$g$ and [M/H]. In Fig.\,3, the
relationships between log~($T_{\rm eff}$) and color index are
shown, and the correlation between log~($T_{\rm eff}$) and color
index is nearly linear. The symbol dots represent a grid of
theoretical models for different metallicity F/G dwarfs. As shown
in Fig.\,3, a simple polynomial can describe the relationship
between log~($T_{\rm eff}$) and color indices. A quadratic
polynomial is as follows:
\begin{equation}
\log~(T_{\rm eff})=A_{0}+A_{1}(CI)+A_{2}(CI)^2,
\end{equation}
where $CI$ is the color index. The coefficients of equation (2)
for different color indices are listed in Table 1. The final
photometric effective temperature can be derived by making the
mean value among the four temperatures from the different color
indices. Chen et al.~(2000) used the most temperature-sensitive
color index in their study to derive the effective temperature.
Our results show that our fitting on $d-o$ is very close to the
red region of their fitting on NGC 288 stars.
\begin{table}[ht]
\begin{center}
\caption{Coefficients of the color index and effective temperature
relation}
\vspace{0.5cm}
\begin{tabular}{ccccc}
\hline \hline
Color index  & $A_{0}$ &  $A_{1}$  &   $A_{2}$   & $\chi^{2}$ \\
\hline
$d-n$ & 3.8474  & $-0.1620$&$-0.0031$&0.011\\
$d-o$ & 3.8471  & $-0.1663$& 0.0158 &0.011\\
$e-o$ & 3.8423  & $-0.1884$& 0.0270 &0.013\\
$e-p$ & 3.8408  & $-0.1873$& 0.0327 &0.013\\
\hline
\end{tabular}
\end{center}
\end{table}

\begin{figure}
\figurenum{3} \centerline{\psfig{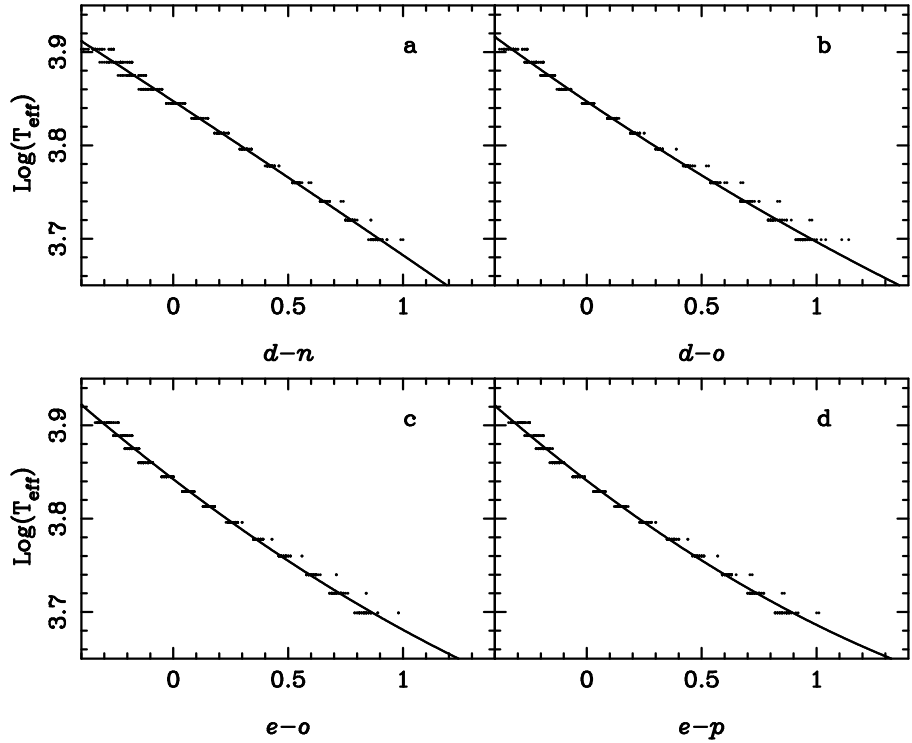}}
\caption{The relationship between log~($T_{\rm eff}$) and the
color indices of bandpasses relatively insensitive to $\log$~g in
our data for (a) $d-n$, (b) $d-o$, (c) $e-o$, (d) $e-p$. The
symbol dots represent a grid of theoretical models for different
metallicity dwarfs.}
\end{figure}

\subsection{Metallicity and abundance}
Photometric indices were often used to give the stellar
metallicity. For example, Wallerstein (1962) derived the [Fe/H]
calibration using the $\delta(U-B)$ excess of the Johnson $UBV$
photometric system. For the $uvby$ system, Crawford (1975a),
Nissen (1981) and Olsen (1988) used the photometric differentials
$\delta~m_{1}$ and  $\delta~c_{1}$ to derive the [Fe/H]
calibration.  Following that, Schuster \& Nissen (1989) made
direct use of the photometric indices $m_1$ and $c_{1}$ to derive
[Fe/H] calibrations for F and G dwarfs.

For the BATC multicolor photometric system, there are 15
intermediate band filters covering an optical wavelength range
from 3000 to 10000 {\AA}. So the SEDs of 15 filters for every
observed object are equivalent to a low resolution spectra. Using
the equation (2), the effective temperature of a given star can be
derived in advance from the available photometry. Thus, given a
photometrically determined effective temperature, we can use the
SED simulation to derive the stellar metallicity. The standard
$\chi^2$ minimization, i.e., computing and minimizing the
deviations between the photometric SED of a star and the template
SEDs obtained with the same photometric system, is used in the
fitting process. The minimum $\chi^2_{min}$ indicates the best fit
to the observed SED by the set of template spectra:
\begin{equation}
{\chi^{2}=\sum\limits_{l=1}^{N_{filt}=15}\left [\frac{m_{obs,l}-
 m_{temp,l}-b} {\sigma_{l}} \right ]^{2}},
\end{equation}
where ${m_{obs,l}}$, $m_{temp,l}$ and $\sigma_{l}$ are the
observed magnitude, template magnitude and their uncertainty in filter
$l$, respectively, and $N_{filt}$ is the total number of filters
in the photometry, while $b$ is the mean magnitude difference
between observed magnitude and template magnitude. The uncertainties of
metallicity obtained from comparing SEDs between photometry and
theoretical models are due to the observational error and the
finite grid of the models (Chen et al.~2000). For the
metal-poor stars ([Fe/H]$<-1.0$), the metallicity uncertainty is
about 0.5 dex, and 0.2 dex for the stars [Fe/H]$>-0.5$.

\section{PHOTOMETRIC DATA}
\subsection{Object classification and photometric parallaxes}
In this study, we use two fields at intermediate latitudes;\\
--- BATC T329 field with central coordinates
${\alpha=09^h53^m13^s{\mbox{}\hspace{-0.13cm}.}30}$ and
$\delta=47^\circ49^{\prime}00^{\prime\prime}{\mbox{}\hspace{-0.15cm}.0}$
(J2000) (Galactic coordinates: $l=169.95^{\circ},
b=49.80^{\circ}$). It is complete to 20.5 mag with an error of
less than $0\hspace{0.1cm}{.}\hspace{-0.15cm}^{m}1$ in the BATC
$i$ band. Each object is classified according to their SED
information constructed from the 15-color photometric catalog. The
observed colors of each object are compared with a color library
of known objects with the same photometric system. Firstly, we use
spectral templates of galaxies and stars to discriminate the stars
from galaxies. The profile of objects classified as stars does not
deviate significantly from that of stellar templates. The input
library for stellar spectra is the Pickles (1998) catalog. Details
about the classification of galaxies are given in Xia et
al.~(2002) and stars given in Du et al.~(2003). The limiting
magnitude is not so deep as to be strongly contaminated by a lot
of galaxies, and the possibility of galaxy contamination
is estimated to be less than 3$\%$. \\
--- BATC TA01 field with central coordinates
${\alpha=01^h12^m06^s{\mbox{}\hspace{-0.13cm}.}00}$ and
$\delta=-00^\circ02^{\prime}00^{\prime\prime}$ (J2000) (Galactic
coordinates: $l=134.16^{\circ}, b=-62.45^{\circ}$). It is complete
to $20\hspace{0.1cm}{.}\hspace{-0.15cm}^{m}0$ in the BATC $i$
band. It should be noted that the stars-galaxies separation for
the BATC TA01 field is different from the BATC T329 field. Because
the BATC TA01 field has been observed by the Sloan Digital Space
Survey ($SDSS$) and each object type (stars-galaxies-QSO) has been
given, we can make direct use of those stars to obtain star types
according to the  stellar spectra library.

According to the results of object classification, we pick out the
F/G dwarfs from the two fields. Details about the classification
of stars can be found in Du et al.~(2003). In total, there are 383
F/G dwarfs, and the photometric parallaxes can be obtained
according to the stellar types. In Fig.~4, we give the actual $z$
distance distribution for two fields. It is clear that most of
stars are in $z\sim1-2$ kpc for the two fields. A variety of
errors affect the determination of stellar distances. The first
source of errors is from photometric uncertainty less than 0.1 mag
in the BATC $i$ band; the second from the misclassification, which
should be small due to the multicolor photometry. For luminosity
class V, types F/G, the absolute magnitude uncertainty is about
0.3 mag. In addition, there may exist an error from the
contamination of binary stars in our sample. \textbf{We neglect
the effect of binary contamination on distance derivation due to
the unknown but small influence from mass distribution in binary
components (Kroupa et al.~1993; Ojha et al.~1996).} The two fields
lie in intermediate latitudes and the influence of interstellar
extinction in the distance calculation can be neglected.

\subsection{Classification comparison and stellar population }

The stellar classifications are compared between stellar spectra
library from Pickles (1998) and theoretical spectra library from
Lejeune (1997). In our previous paper (Du et al.~2003), we derived
the stellar type according to Pickles stellar spectra library
(Pickles 1998). To obtain more information about metallicity, we
here use theoretical spectra library from Lejeune (1997) to derive
the stellar parameters such as temperature and metallicity. In
order to check the agreement between the Pickles library and
Lejeune library, we present a comparison.

As an example, Table 2 lists the comparison results between
Pickles spectra and Lejeune spectra for six stars in our sample.
The first column is the stellar type derived from Pickles library.
According to the theoretical library from Lejeune (1997), the
effective temperature $T_{\rm eff}$ is determined by using
equation (2), and it is listed in the second column. The gravity
$\log~g$ and metallicities ([Fe/H]) are also derived by fitting
SEDs according to the theoretical library from Lejeune (1997).
They are listed in the following two columns. The apparent visual
magnitude $V$ and photometric distances are listed in the last two
columns. \textbf{As listed in Table 2, our classification results
from Pickles library (1998) are in excellent agreement with
theoretical library from Lejeune (1997).} The metallicities
([Fe/H]) for these stars can only provide an estimate of
individual stellar abundances, but we have the advantage of being
able to use many stars to obtain mean metallicity at different
distances from the plane.

\begin{table}[ht]
\begin{center}
\caption{Classification comparisons between stellar synthetic
spectra from Lejeune (1997) and stellar library from Pickles
(1998)}
\vspace {0.5cm}
\begin{tabular}{cccccc}
\hline \hline
type&$T_{\rm eff}$ &[Fe/H] &$\log~g$ & $V$ & $r$ (kpc)\\
\hline
 G8V & 5110& $-2.0$& 4.5& 18.2& 2.9\\
 G5V & 5330& $-1.0$& 4.5& 17.2& 2.4\\
 G3V & 5470& $-0.5$& 4.0& 16.1& 1.8\\
 G0V & 5717& $-1.0$& 4.5& 16.5& 2.4\\
 F7V & 6047& $-0.3$& 4.5& 14.6& 1.5\\
 F5V & 6273& $-3.5$& 4.0& 17.0& 4.0\\
\hline
\end{tabular}
\end{center}
\end{table}

Standard star-count models indicate that the color-magnitude range
could be used to separate roughly different populations of the
Galaxy (Chen et al.~2001, Du et al.~2003). The present sample
stars have colors and magnitudes in the ranges $0.1<d-i<0.9$ and
$14.0<i<20.5$. According to star-count models, the present sample
should contain predominantly thick disk stars in the
color-magnitude range, with some contributions from the thin disk
and halo near the edge of the sample selection. As shown in
Fig.~5, there are no stars redder than $(d-i)\sim0.9$. For the
T329 field, the stars distribution is smooth, while there is a
sharp increase at $(d-i)\sim0.6$ for the TA01 field. The peak
values for the two fields, however, lie in $(d-i)\sim0.6$. We use
the Galaxy models to predict the relative frequency of dwarfs
belonging to each of the three dominant components in the Galaxy.
These relative frequencies obviously depend on both the local
normalization of each component's density distribution and its
scale height. The density distributions can be combined with the
local volume element to produce estimates of the number of stars
from given components expected to be observed at a given height
above the plane (Gilmore \& Wyse 1985),
\begin{equation}
n_{i}(z)=\rho_{0i}~\exp{(-z/z_{0i})}~z^{2}{\rm d}z,
\end{equation}
where the subscript $i$ refers to each component, $\rho_{0}$ being
the local density normalization, and $z_{0}$ the scale height. The
components are (i) the thin disk with local normalization 1.0 and
an exponential scale height of 320 pc (Du et al.~2003); (ii) the
thick disk, with local normalization 0.07 and exponential scale
height of 640 pc (Du et al.~2003); and (iii) the halo, with local
normalization 0.00125 and $R^{1/4}$ density distribution with
axial ratio of 0.6, and an effective radius of 2.7 kpc (Du et al.
2003). The resultant curves for various components of the Galaxy
are shown in Fig.~6. The solid line, dotted-dashed line and dotted
line represent the contribution of the halo, thick disk and thin
disk, respectively. The various curves are normalized
independently of each other. In Table 3, we show the logarithmic
numbers as functions of height. The number of stars in the thin
disk, thick disk, and halo are represented by $n_{0}$, $n_{1}$ and
$n_{2}$ respectively. The curves in Fig.~6, when combined with
local normalization, mean that the thin disk will contribute up to
$z$ heights of about 1 kpc, while the thick disk will dominate
from 1 kpc to 4 kpc.

\begin{table}[ht]
\begin{center}
\caption{The number of stars versus the distance $z$ for three components of the Galaxy }
\vspace{0.5cm}
\begin{tabular}{cccc}
\hline \hline
$z$ (kpc)  & log ($n_{0}$) &  log ($n_{1}$)  &  log  ($n_{2}$)    \\
\hline
  0.05&  4.03&  2.91&  1.19\\
  0.69&  5.44&  4.76&  3.36\\
  1.34&  5.13&  4.89&  3.81\\
  1.98&  4.60&  4.79&  4.03\\
  2.63&  3.97&  4.60&  4.14\\
  3.27&  3.28&  4.35&  4.20\\
  3.92&  2.57&  4.07&  4.23\\
  4.57&  1.82&  3.77&  4.24\\
  5.21&  1.06&  3.44&  4.24\\
  5.86&  0.29&  3.11&  4.23\\
\hline
\end{tabular}
\end{center}
\end{table}


\begin{figure}
\figurenum{4}
\centerline{\psfig{file=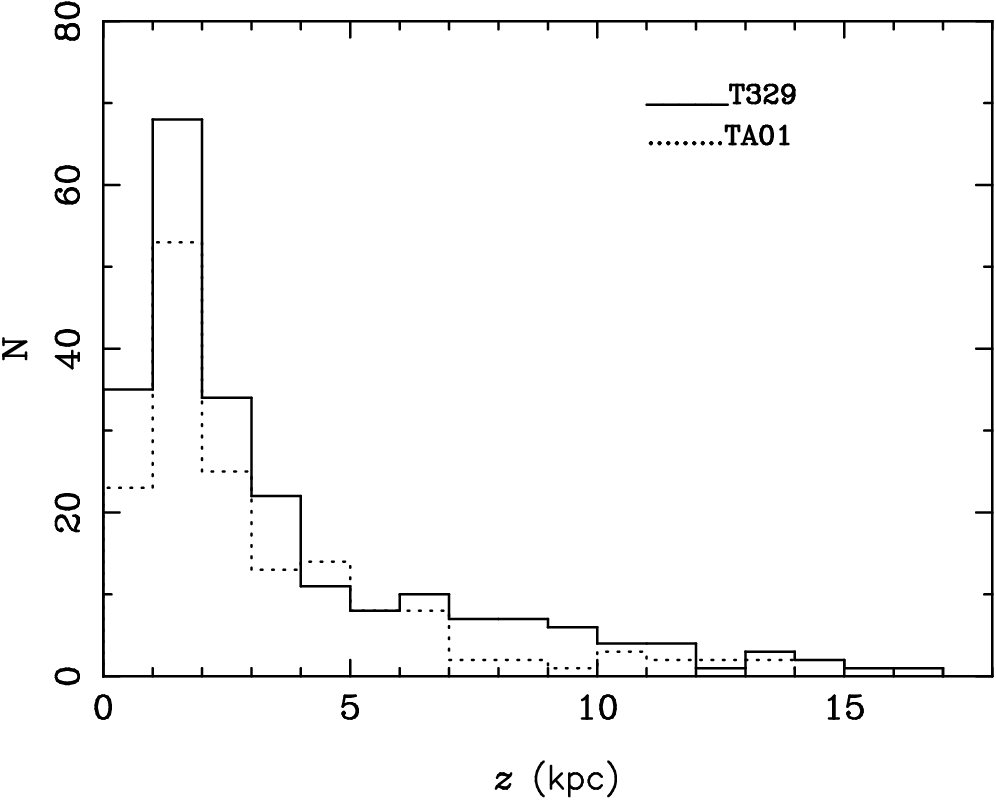,width=10.0cm}}
\caption{Distribution of $z$ distance in two fields. The solid
line represents the T329 field and the dotted line represents the
TA01 field. }
\end{figure}


\begin{figure}
\figurenum{5} \centerline{\psfig{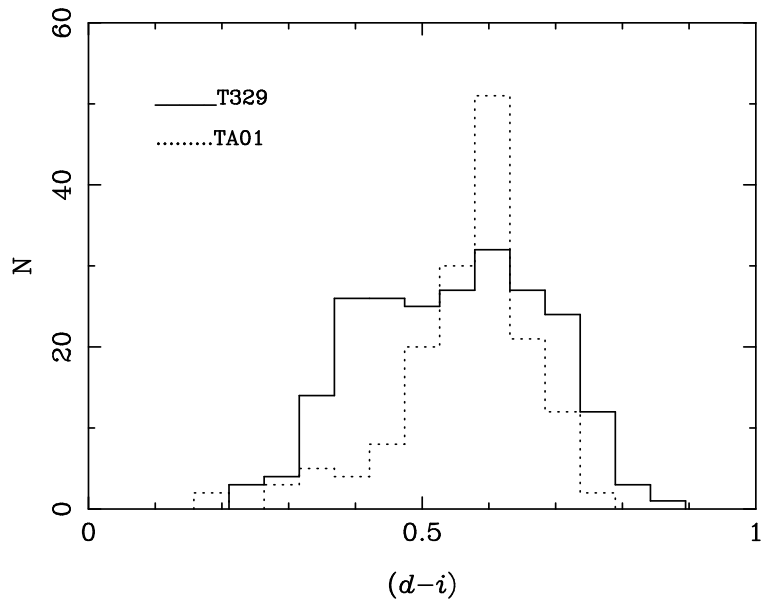}}
\caption{Distribution of $(d-i)$ in two fields. The solid line
represents the T329 field and the dotted line represents the TA01
field.}
\end{figure}

\begin{figure}
\figurenum{6} \centerline{\psfig{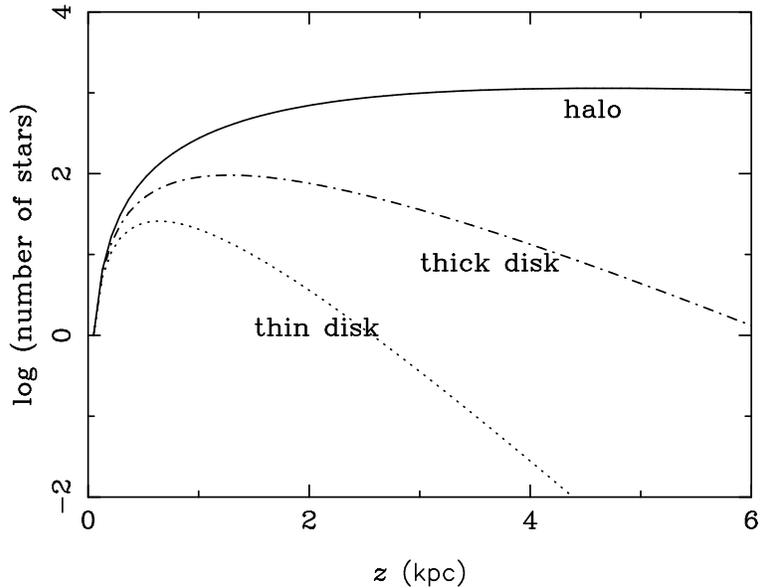}}
\caption{Relative number of stars within a given volume element as
a function of height above the plane for various components of the
Galaxy.}
\end{figure}

\section{THE VERTICAL METALLICITY GRADIENT TOWARDS HIGH LATITUDE
FIELDS}

It is well known that the chemical abundance of a stellar
population contains much information about the population's early
evolution, while detailed information about the vertical
metallicity gradient can provide an important clue about the
formation scenario of stellar populations. The metallicity
distribution of stars in the Galaxy has been the subject of
several spectroscopic and photometric surveys (Yoss et al.~1979;
Hartkopf et al.~1982; Gilmore et al.~1985; Ratnatunga et al.~1989;
Friel~1988).

The existence of a vertical metallicity gradient among field and
open cluster stars of the Galactic disk is controversial. For
example, using DDO photometry of the late-type giants in the
direction of both galactic poles, several authors found a vertical
gradient ranging from approximately $-0.2$ to $-0.4$ dex
kpc$^{-1}$ (see Hartkopf \& Yoss 1982; Yoss et al. 1987; Norris \&
Green 1989). Other independent studies have also shown evidence
for a significant vertical gradient: Buser \& Rong (1995b), using
the photographic RGU data to study the metallicity distribution of
the Galactic components, found a vertical abundance gradient of
$-0.6$ dex/kpc for the old thin disk and a marginal metallicity
gradient of $-0.1$ dex/kpc for the thick disk. Robin et al. (1996)
showed that the data towards the pole are in favor of a small
gradient of $-0.25$ dex kpc$^{-1}$, data in SA54 field (Yamagata
\& Yoshii 1992) give a high value of $-0.65$, while data from
Fenkart \& Esin-Yilmaz (1985) favor a small gradient of $-0.15$
dex kpc$^{-1}$. However, Gilmore (1985) showed that there is
little or no gradient in the three components of the Galaxy.

Recent work has used open clusters instead of field stars to
determine the metallicity gradient perpendicular to the disk.
Piatti et al.~(1995) found a correlation between [Fe/H] and
vertical position and they obtained a gradient equal to $-0.34$
dex kpc$^{-1}$ by binning the data in $z$. Carraro (1998) also
derived a vertical abundance gradient of $-0.25$ dex kpc$^{-1}$.
Cameron (1985) used metallicities determined from $UBV$ photometry
and found no gradient perpendicular to the plane. Friel (1995) did
not find a gradient from the study of open clusters either.

\begin{figure}
\figurenum{7} \centerline{\psfig{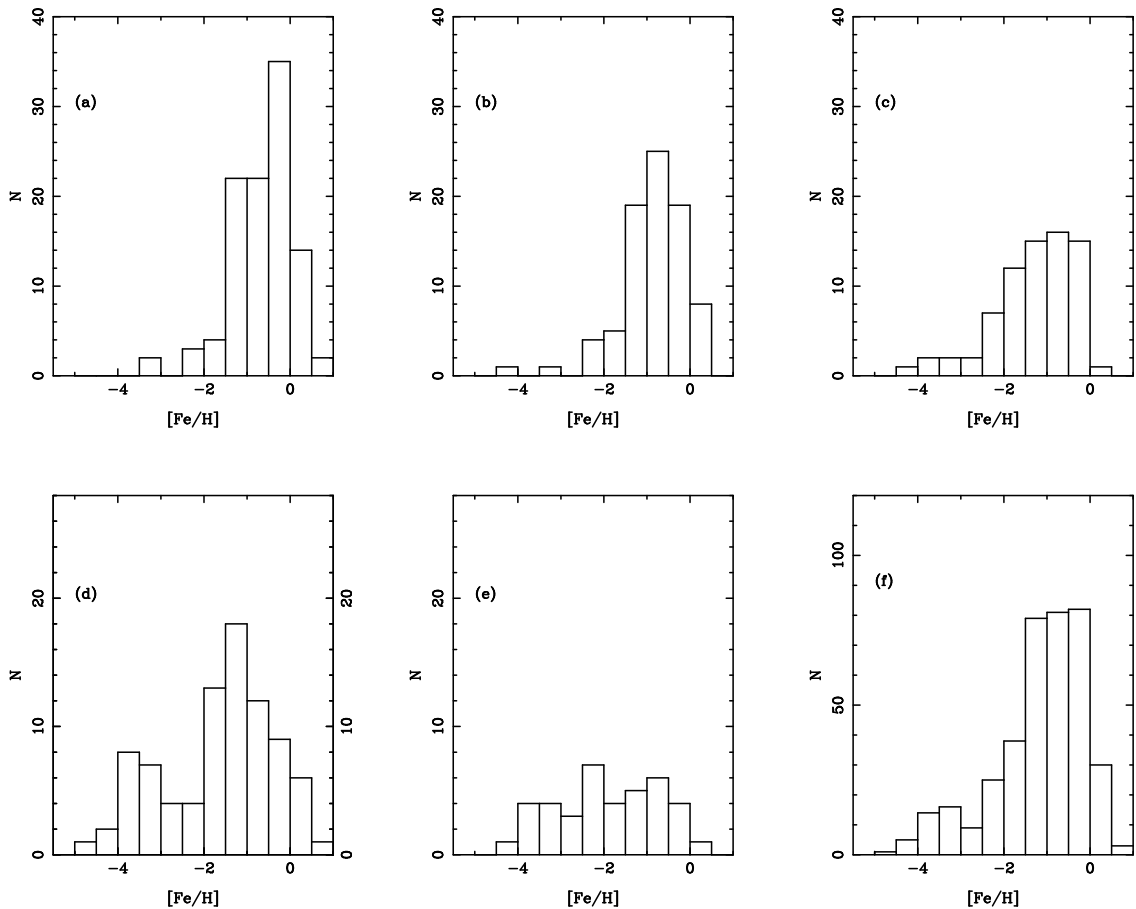}}
\caption{Metallicity distribution for sample stars  as functions
of apparent magnitude $i$. (a) $14.0<i\le15.5$, (b)
$15.5<i\le16.5$, (c) $16.5<i\le17.5$, (d) $17.5<i\le18.5$, (e)
$i>18.5$ and (f) $14.0<i\le20.5$.}
\end{figure}

In this study, we want to show how the BATC survey data can limit
possible metallicity gradients for the components of the Galaxy.
At first, the metallicity for the sample F/G dwarfs can be derived
by comparing SEDs between photometry and theoretical models. The
SEDs fitting method is already described in Sect 3.3. In
Fig.\,7a-e, the metallicity distributions are shown as functions
of apparent magnitude for sample stars. Fig.\,7f gives the
metallicity distribution for all stars. Here, we divide these F/G
dwarfs into 5 bins: $14.0<i\le15.5$, $15.5<i\le16.5$,
$16.5<i\le17.5$, $17.5<i\le18.5$ and $i>18.5$ to see how the
magnitude affects the abundance distribution. From the figures
(Fig. \,7a-e), we can see that there is a shift from metal-rich
stars to metal-poor ones with the increasing of apparent
magnitude. It is particularly apparent in Fig.\,8, where the mean
metallicity as a function of $z$-distance is displayed for the
combined sample. It should be noted that we use the mean
metallicity to describe the metallicity distribution function. We
adopt the `biweight' statistic method provided by ROSTAT software
(Beers et al.~1990) to estimate the stellar mean metallicity.
Although computationally more complex, the `biweight' estimate has
proved to be superior in many respects and has also performed well
for small samples. Karaali et al.~(2003) showed that the mean
value is also valid to describe a distribution function.

As Fig.\,8 shows clearly, the mean metal abundance decreases with
the increasing mean $z$, indicating a clear vertical metallicity
gradient of the disk ($z<4$ kpc). The individual fields give
essentially the same result, and therefore have been combined to
improve the statistics. The T329 field lies in north latitude
direction and TA01 field in south latitude direction. In addition,
the distributions in the two fields do not differ significantly in
number of stars, so we can combine them.  The overall distribution
shows a metallicity gradient d[Fe/H]/d$z=-0.17\pm0.04$~dex/kpc, up
to 15 kpc. At the same time, we find a gradient of
$-0.37\pm0.1$~dex/kpc for $z<4$ kpc, whereas it shows a weak or
zero gradient between 5 and 15 kpc, i.e.,
d[Fe/H]/d$z=-0.06\pm0.09$~dex/kpc.

Based on Geneva photometry, Grenon (1977) investigated the
vertical metallicity gradient for G and K giants and estimated a
value of $-0.35$ dex/kpc for $z$ between 0 and 700 pc. Yoss et
al.~(1987) obtained G5-K6 giants near the galactic poles and found
a chemical gradient of $-0.4$ dex/kpc for the thin disk and a
gradient of $-0.18$ dex/kpc extending to $z=8$ kpc, a value that
is identical with the one found in this study for $z<15$ kpc.
Yoshii et al.~(1987) derived a metallicity gradient of
$-0.5\pm0.1$ dex/kpc for $z\leq2$ kpc in logarithmic from UBV star
count data by assuming the solar abundance in the Galactic plane,
while this value is slightly steeper than the value of $-0.37$
dex/kpc for $z<4$ kpc in this study. J\"{o}nch-S\"{o}rensen (1994)
observed a sample of faint F-dwarfs in order to investigate the
spatial distribution of metallicity in the disk of the Galaxy.
They found a vertical gradient of $-0.33$ dex/kpc for $z\le500$ pc
and $-0.19$ dex/kpc for 500 $\le z\le$1500 pc. Trefzger et
al.~(1995) found a rather steep mean metallicity gradient of
$-0.55\pm0.1$ dex/kpc up to $z=900$ pc and determined an overall
gradient of $-0.23$ dex/kpc for $z<4$ kpc, this gradient is
slightly flatter than the one found in this study. We noted that
these results are in close agreement, both quantitatively and
qualitatively, with our finding at the high latitude fields. In
Sect.~6, our results are similarly interpreted as a mixture of
stellar population with different mean metallicities at all $z$
levels.

\begin{figure}
\figurenum{8} \centerline{\psfig{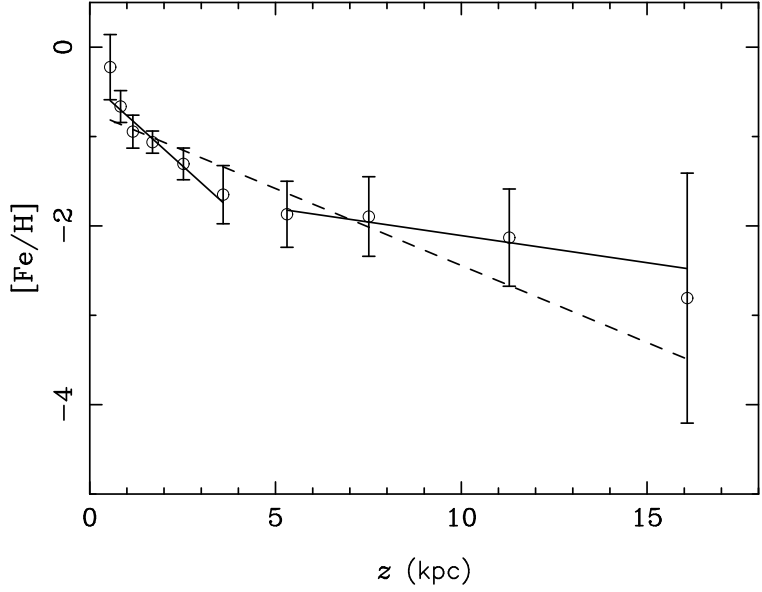}}
\caption{Mean metallicity distribution versus mean $z$-distance
(kpc) for 10 $z$-intervals, showing a metallicity gradient
d[Fe/H/d$z\sim-0.37$~dex/kpc for $z<4$ kpc  and
d[Fe/H/d$z\sim-0.06$~dex/kpc(or possibly zero) between 5 and 15
kpc}
\end{figure}

\section{THE METALLICITY GRADIENT ARISING FROM THE POPULATION
GRADIENT}

A crucial aid to the interpretation of these metallicity gradients
is the varying contributions from the different Galactic
components. As shown in Fig.~6, the expected relative number of
stars from different components of the Galaxy is a function of
height from the Galactic plane. The components are the thin disk,
thick disk and halo. The thin disk population contains a young
(age $\le3$ Gyr), metal-rich population (mean abundance
$<$[Fe/H]$>\sim0.0$; $\sigma_{[Fe/H]}\sim0.15$), and a older (age
$\ge$3 Gyr) more metal-poor population (mean abundance
$<$[Fe/H]$>\sim-0.3$; $\sigma_{[Fe/H]}\sim0.2$) (Gilmore \& Wyse
1985). The younger population contains $\sim 20\%$ of the stars
near the sun and has a vertical scale height $\sim$ 100 pc. The
older population contains $\sim 80\%$ of the stars near the sun,
and is characterized by a exponential scale height of $\sim$ 300
pc. The young thin disk stars contribute up to $z$ heights of
around 300 pc and these old thin disk stars dominate from 300 pc
to 1 kpc. Here, we only consider the old thin disk population in
our model since we have very few samples of very nearby stars.

For the thick disk, Hartkopf \& Yoss (1982) derived the mean
abundance $<$[Fe/H]$>\sim-0.6$ and  $\sigma_{[Fe/H]}\sim0.3$ with
distances between 1 and 2 kpc; Buser (1999) also derived the mean
metallicity of the thick disk $<$[Fe/H]$>\sim-0.63$ and dispersion
$\sigma_{[Fe/H]}\sim0.4$ dex. Chiba \& Beers (2000) showed that
the thick disk population includes stars with a wide range of
metallicity, from $-2.2\le$[Fe/H]$\le-0.5$, and  most of stars are
in the more-rich end of this range. Recently, based on
medium-resolution spectroscopy and broadband photometry, Beers et
al.~(2002) found that the local fraction of metal-poor stars which
might be associated with the metal-weak thick disk is on the order
of $30\%-40\%$ at abundances below [Fe/H]=$-1.0$. At the same
time, they also found that this relatively high fraction of local
metal-poor stars may extend to metallicities below [Fe/H]=$-1.6$,
much lower than what had been considered before. For the halo, the
field stars have similar mean metallicity with globular clusters
in the Milky Way. However, the halo field stars extend to much
lower metallicity ([Fe/H]$\simeq-5$) than that of the globular
clusters ([Fe/H]$\simeq-2.2$) (Freeman et al.~2002). In general,
it is adopted here that the halo mean abundance
$<$[Fe/H]$>\sim-1.5$ and $\sigma_{[Fe/H]}\sim0.5$ (Gilmore \& Wyse
1985).

In this study, we try to estimate the metallicity gradient
resulting from changing relative proportions of different
populations, assuming there doesn't exist a gradient for a single
population. The number of stars at different heights $z$ for three
components (Fig.~6) could be combined with mean metallicity to
produce estimates of metallicity distributions as a function of
height above the plane. In order to check the mean metallicity's
effect on the gradient, we divide the mean metallicities into two
cases to discuss. In case A, the mean metallicities are
$<$[Fe/H]$>\sim0.0$ for the thin disk, $<$[Fe/H]$>\sim-0.6$ for
the thick disk, and $<$[Fe/H]$>\sim-1.5$ for the halo,
respectively. In case B, the mean metallicities are,
$<$[Fe/H]$>\sim-0.3$ for the thin disk, $<$[Fe/H]$>\sim-1.0$ for
the thick disk, and $<$[Fe/H]$>\sim-2.0$ for the halo,
respectively. The mean parameter values in case A are in agreement
with the majority of recent determinations, while the values in
case B lie in the range of error of case A. The results are
illustrated in Fig.~9; the dotted line and solid line represent
case A and case B, respectively. The open circles represent the
observational data. Many factors affect the determination of mean
metallicity. The first source contributing to errors is from
photometric errors; the second from stellar metallicity by fitting
SEDs. In addition, there may exist an error from incompleteness of
samples in each bin. As Fig.~9 shows clearly, although case B
matches the observed metallicity distribution, both cases exhibit
similar gradients for $z<5$ kpc and they become flat for $z>5$
kpc. Thus the gradient we derived ($-0.37\pm0.1$ dex/kpc for $z<4$
kpc) can be interpreted as the different contributions in density
distribution for the three components of the Galactic model,
whereas the small or zero gradient d[Fe/H]/d$z=-0.06\pm0.09$ can
not be interpreted due to the different contributions from three
components. In particular the last datum point ($z\sim15$ kpc) can
be attributed to the incompleteness of the sample at that large
distance. So it is possible that there is a little or no gradient
for $z>5$ kpc. It should be noted that the low mean metallicity in
case B is presumably a consequence of the selection effects (such
as the magnitude, color selection and distance selection, etc.)
involved in the definition of the sample. But this is an
unimportant effect for the derivation of the gradient.

\begin{figure}
\figurenum{9} \centerline{\psfig{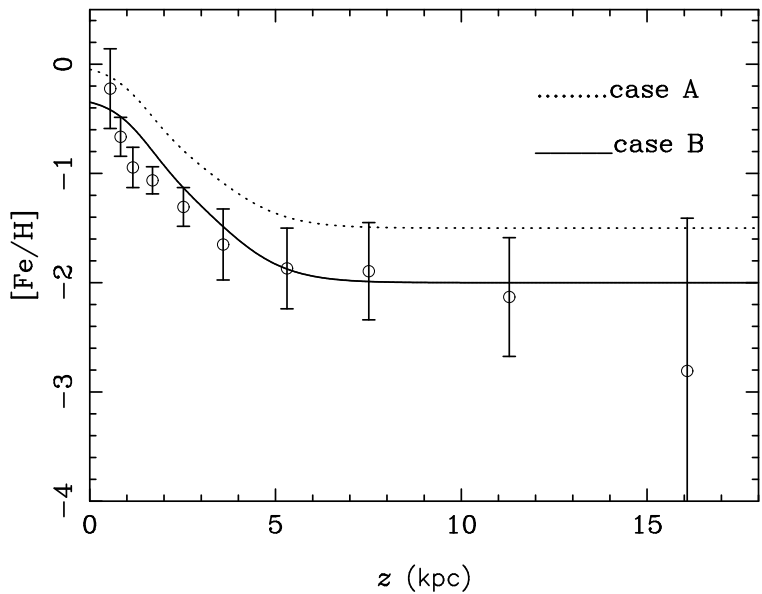}}
\caption{Observed and calculated mean metallicity distribution as
a function of height $z$. The results of case A (dotted line) and
case B (solid line) are illustrated.}
\end{figure}

\section{CONCLUSIONS AND SUMMARY}

In this work, based on the BATC multicolor photometric system, we
develop a calibration of temperature and metallicity for the
dwarfs by employing a set of synthetic flux spectra calculated
from theoretical models. Some temperature-sensitive color indices
in the BATC filter system are found to yield well-defined liner
relations with log~($T_{\rm eff}$) (see Fig.~3). \textbf{More
importantly, this study shows that in using the BATC multicolor
observations not only can we conveniently assess stellar
temperature, but metallicity and gravity as well.} We thus have
confirmed the fact that stars can be successfully classified
three-dimensionally from the BATC SEDs.

In addition, in this paper, by using F/G dwarfs of two BATC
fields, we determined the average stellar metallicity as a
function of distance from the galactic plane. It can clearly be
seen that the mean metallicity decreases with increasing $z$,
namely, we find a gradient of $-0.37\pm0.1$~dex/kpc for $z<4$ kpc,
whereas it shows a weak gradient between 5 and 15 kpc, i.e.,
d[Fe/H]/d$z=-0.06\pm0.09$~dex/kpc. The overall distribution shows
a metallicity gradient d[Fe/H]/d$z=-0.17\pm0.04$~dex/kpc, up to 15
kpc. These results are in agreement with the values in the
literature (Yoshii et al.~1987; Yoss et al.~1987; Trefzger et
al.~1995). In our study, these results are interpreted as
different contributions from three components of the Galaxy at
different $z$ distances. In star counts the younger metal-rich
stars are confined to regions close to the Galactic mid-plane,
while the older, metal-poorer stars with a larger scale height
dominate at larger vertical distances from the Galactic plane. As
a consequence, a vertical metallicity gradient is caused by the
varying dominance of different stellar components of the Galaxy.
It is possible that additional observational investigations will
give more evidence for the metallicity gradient of the Galaxy and
therefore provide a powerful clue to the disk and halo formation
because it almost certainly concerns events during and shortly
after the early Galactic collapse and the beginning of disk
formation (Sandage 1981; Bell 1996; Norris \& Ryan 1991).

\acknowledgments We would like to thank the referee, Dr. Yoss, for
his insightful comments and suggestions that improved this paper
greatly. The BATC Survey is supported by the Chinese Academy of
Sciences, the Chinese National Natural Science Foundation under
the contract No. 10273012 and the Chinese State Committee of
Sciences and Technology. This work has been supported by the
National Key Basic Research Science Foundation (NKBRSF
TG199075402). We also thank the assistants who helped with the
observations for their hard work and kind cooperation.

\end{document}